\documentclass[12pt]{article}

\usepackage{times}
\usepackage{epsfig}
\usepackage{hyperref}
\usepackage{amssymb}
\usepackage{amsmath}
\input{epsf}
\newcommand\as{\alpha_{\mathrm{S}}}

\def\beq{\begin{equation}}
\def\eeq{\end{equation}}
\def\beeq{\begin{eqnarray}}
\def\eeeq{\end{eqnarray}}

\def\to{\rightarrow}

\newcommand{\la}{\langle}
\newcommand{\ra}{\rangle}

\def\nn{\nonumber}

\def\ID{1 \kern -.45 em 1}

\def\ket#1{|{#1}\ra}
\def\bra#1{\la{#1}|}

\begin{document}

\begin{titlepage}
\renewcommand{\thefootnote}{\fnsymbol{footnote}}
\begin{flushright}
ICAS 39/18 \\
     \end{flushright}
\par \vspace{10mm}
\begin{center}
{\large \bf
The Spin Budget of the Proton at NNLO and Beyond}\\

\vspace{8mm}

\today
\end{center}

\par \vspace{2mm}
\begin{center}
{\bf Daniel de Florian${}^{\,a}$,}
\hskip .2cm
{\bf Werner Vogelsang${}^{\,b}$  }\\[2mm]
\vspace{5mm}
${}^{a}\,$ International Center for Advanced Studies (ICAS), ECyT-UNSAM,
Campus Miguelete, \\ 25 de Mayo y Francia, (1650) Buenos Aires, Argentina\\[2mm]
${}^{b}\,$ Institute for Theoretical Physics, T\"ubingen University, 
Auf der Morgenstelle 14, \\ 72076 T\"ubingen, Germany\\
\end{center}

%%%%%%%%%%%%%%%%%%%%%%%%%%%%%%%%%%%%%%%%%%%%%%%%%%%%%%%%
%%%%%%%%%%%%%%%%%%%%%%%%%%%       ABSTRACT      %%%%%%%%%%%%%%%%%%%
%%%%%%%%%%%%%%%%%%%%%%%%%%%%%%%%%%%%%%%%%%%%%%%%%%%%%%%%

\vspace{9mm}
\begin{center} {\large \bf Abstract} \end{center}
We revisit the scale evolution of the quark and gluon spin contributions to the proton spin,
$\frac{1}{2}\Delta \Sigma$ and $\Delta G$, using the three-loop results for the 
spin-dependent evolution kernels available in the literature. We argue that the evolution of the 
quark spin contribution may actually be extended to four-loop order, and that to 
all orders a single anomalous dimension governs the evolution of both $\Delta \Sigma$ and $\Delta G$.
We present analytical solutions of the evolution equations for $\Delta \Sigma$ and $\Delta G$
and investigate their scale dependence both to large and down to lower ``hadronic'' scales.
We find that the solutions remain perturbatively stable even to low scales, where they
come closer to simple quark model expectations.
We discuss a curious scenario for the proton spin, in which even the gluon spin contribution is 
essentially scale independent and has a finite asymptotic value as the scale becomes large. 
We finally also show that perturbative three-loop evolution leads to a larger spin contribution 
of strange anti-quarks than of strange quarks. 

\end{titlepage}  

\renewcommand{\thefootnote}{\fnsymbol{footnote}}

\section{Introduction}

The decomposition of the proton spin in terms of the contributions by quarks and anti-quarks,
gluons, and orbital motion is a key focus of modern nuclear and particle physics. As has become
well-known~\cite{Jaffe:1989jz,Ji:1996ek,Leader:2013jra}, in a gauge theory the decomposition is
not unique. The two physically most relevant spin sum rules for the proton are the Ji decomposition~\cite{Ji:1996ek},
which ascribes the proton spin to gauge-invariant contributions by quark spins and orbital 
angular momenta, and total gluon angular momentum, and the Jaffe-Manohar decomposition~\cite{Jaffe:1989jz},
in which there are four separate pieces corresponding to quark and gluon spin and orbital contributions,
respectively. The two sum rules have in common only the quark spin piece, 
%WV added part of your sentences below. We can use them or drop them (the referee will not 
%care). Either way is fine with me.
and there is no relation among
the other pieces. In particular, in the gauge-invariant definition of Ref.~\cite{Ji:1996ek} only the 
total gluon angular momentum is well defined and cannot be split in a physically meaningful 
way into helicity and orbital angular momentum contributions.

The Jaffe-Manohar sum rule corresponds to the canonical decomposition of 
the proton's angular momentum. It may be regarded as a ``partonic'' spin sum rule, since both the
quark and gluon spin pieces are related to parton distributions measurable in inelastic $\ell p$ or $pp$ 
scattering processes. The sum rule reads
\beq\label{JM}
\frac{1}{2}\,=\,\frac{1}{2}\Delta \Sigma(Q^2)+\Delta G(Q^2)+L_q(Q^2)+L_g(Q^2)\,,
\eeq
where $\frac{1}{2}\Delta \Sigma$ and $\Delta G$ are the quark and gluon spin contributions 
and $L_q$ and $L_g$ the orbital ones. $\Delta \Sigma$ and $\Delta G$ may be obtained from the 
first moments of the helicity parton distributions $\Delta q(x,Q^2)$, $\Delta \bar{q}(x,Q^2)$
(where $q=u,d,s,\ldots$) and $\Delta g(x,Q^2)$ of the proton:
\beeq\label{singsec}
\Delta \Sigma(Q^2)&=& \sum_q^{N_f} \int_0^1 dx \,
\Big( \Delta q(x,Q^2)+ \Delta \bar{q}(x,Q^2)\Big)\,,\nn\\[2mm]
\Delta G(Q^2) &=&\int_0^1 dx \,\Delta g(x,Q^2)\,,
\eeeq
where in the first line the sum runs over all active quark flavors whose number we denote by $N_f$. 
In the following we will mostly use the simplified notation 
\beq
\Delta q(Q^2)\,\equiv\, \int_0^1 \,dx \, \Delta q(x,Q^2) \; ,
\eeq
and likewise for the anti-quarks. 
%WV new sentences. Not sure about the second one...
We note that although the parton distribution $\Delta g(x,Q^2)$ and hence its first
moment are gauge-invariant, the identification with the gluon spin contribution
is only valid in the light-cone gauge. The same is true for the orbital angular
momentum pieces. 

As indicated in Eq.~(\ref{JM}), the contributions to the proton spin are all scale dependent,
although the dependence cancels in their sum. The dependence on $Q^2$ is given by 
spin-dependent QCD evolution equations. The kernels relevant for the evolution of the first 
moments $\Delta q(Q^2)$, $\Delta \bar{q}(Q^2)$, $\Delta G(Q^2)$ have been derived 
to lowest order (LO) in Refs.~\cite{Ahmed:1976ee,Altarelli:1977zs}, to next-to-leading order (NLO) 
in Refs.~\cite{Kodaira:1979pa,Mertig:1995ny,Vogelsang:1995vh,Vogelsang:1996im}, 
and recently to next-to-next-to-leading order (NNLO) in 
Refs.~\cite{Vogt:2008yw,Moch:2014sna,Moch:2015usa}. The kernels for the separate evolution 
%WV added "in the JM decomposition
of $L_q$ and $L_g$ in the Jaffe-Manohar decomposition 
are known only to LO~\cite{Ji:1995cu,Hagler:1998kg,Harindranath:1998ve},
although the evolution of their sum is known from Eq.~(\ref{JM}) to the same order as that of 
$\frac{1}{2}\Delta\Sigma+\Delta G$, that is, to NNLO. 

In this letter, we will discuss some features of the higher-order scale evolution of the first 
moments, starting from the NNLO results of~\cite{Vogt:2008yw,Moch:2014sna,Moch:2015usa}
for the evolution kernels. We will first study the singlet evolution, where we will extend previous 
arguments by Altarelli and Lampe~\cite{Altarelli:1990jp} to show
that the evolution of $\Delta\Sigma$ may actually be determined even to 
next-to-next-to-next-to-leading order (N$^3$LO). The evolution equations for 
$\Delta\Sigma$ and $\Delta G$ may straightforwardly be decoupled and solved
in closed form. We present the solutions in analytical form and show numerical
results for their evolution at various perturbative orders. We note that 
studies along these lines were first presented to LO in Refs.~\cite{Ratcliffe:1987dp,Ramsey:1988dm,Stratmann:2007hp,Thomas:2008ga}.
The paper~\cite{Altenbuchinger:2010sz} considered the NNLO evolution of $\Delta\Sigma$.
In our paper we go beyond the previous work by extending the results for the evolution of $\Delta\Sigma$ 
to N$^3$LO and that of $\Delta G$ to NNLO. Apart from the intrinsic value of this, we believe 
that our results could also have interesting applications in comparisons to models and 
in studies of nucleon spin structure in lattice QCD~\cite{Lin:2017snn}: 
Although nowadays renormalization on the lattice is typically performed at nonperturbative level, comparison 
to high-order perturbative evolution should be valuable as a benchmark. 

As is well-known~\cite{Altarelli:1990jp,Altarelli:1988nr,Altarelli:1988mu} the gluon spin contribution
$\Delta G(Q^2)$ in general evolves as the inverse of the strong coupling constant and thus rises
logarithmically with growing scale, either to large positive or negative values, depending on the
input $\Delta G(Q_0^2),\Delta\Sigma(Q_0^2)$ at some scale $Q_0$. However, in between there is a unique solution for 
which the gluon spin contribution remains almost flat in $Q^2$ and tends to a finite asymptotic value. 
Such a ``static'' solution in fact occurs~\cite{wvtalk} in the early NLO DSSV 
analysis~\cite{deFlorian:2009vb,deFlorian:2008mr}. We show that static solutions may be
found at every order in perturbation theory and determine the asymptotic values for the gluon
spin contribution at LO, NLO, and NNLO. 

We will finally also study the evolution in the flavor non-singlet sector. Higher-order
evolution is known to generate interesting patterns of flavor- or charge-symmetry breaking
in the nucleon sea. It was shown a long time ago~\cite{ross,Furmanski:1981cw,Stratmann:1995fn} that
NLO evolution leads to an asymmetry $\bar{u}\neq\bar{d}$ both in the unpolarized and
the helicity parton distributions. At NNLO, a new type of valence splitting function 
emerges~\cite{Moch:2004pa,Moch:2015usa,Catani:1994sq}, which gives rise to a difference in the 
strange and anti-strange parton distributions in the nucleon~\cite{Catani:2004nc}, 
just from the fact that the nucleon carries net up and down valence distributions. 
In Ref.~\cite{Catani:2004nc} estimates for the spin-averaged $s(x,Q^2)-\bar{s}(x,Q^2)$ 
were given that showed that the asymmetry resulting from evolution is not as small as might
be expected from a three-loop effect. Of course, non-perturbative physics may well 
be the dominant source of the strangeness asymmetry in the nucleon~\cite{Wang:2016ndh}.
In the present paper we will extend the perturbative study in~\cite{Catani:2004nc} to the 
spin-dependent case. An interesting difference with respect to the spin-averaged asymmetry 
is that the first moment $\int_0^1 dx\,\big(\Delta s(x,Q^2)-\Delta \bar{s}(x,Q^2)\big)$ does
not have to vanish, whereas $\int_0^1 dx\,\big(s(x,Q^2)-\bar{s}(x,Q^2)\big)=0$ due to the
fact that the nucleon does not carry net strangeness. As a result, strange quarks and
anti-quarks may make different contributions to the proton spin. Indeed, as will be a 
result of this paper, such a net strangeness helicity asymmetry arises from NNLO 
evolution.

\section{Evolution equations}

We start by considering the generic evolution equation
for the first moment of a spin-dependent parton distribution $a,b\equiv u,\bar{u},d,\bar{d},s,\bar{s},\ldots,G$:
\beq \label{evN}
\frac{d\Delta a(Q^2)}{d\ln Q^2} = \sum_b \,
\Delta P_{ab}\big(a_s(Q^2)\big) \;\Delta b(Q^2) \;,
\eeq
where $\Delta P_{ab}$ describes the splitting $b\to a$ (it is the first moment
of the usual $x$-dependent splitting function). The $\Delta P_{ab}$ are perturbative 
in the strong coupling $\as$; their perturbative series starts at ${\cal O}(\as)$:
\beq
\label{eq:split}
\Delta P_{ab}=a_s\Delta P_{ab}^{(0)}+
a_s^2 \Delta P_{ab}^{(1)}+
a_s^3 \Delta P_{ab}^{(2)}+ {\cal O}\big(a_s^4\big)\, .
\eeq
with $a_s\equiv \as/(4\pi)$. The running coupling obeys the renormalization group
equation
\beq
\label{eq:alpha}
\frac{d\ln a_s(Q^2)}{d\ln Q^2} \equiv \frac{\beta(a_s)}{a_s} = -\beta_0 a_s -\beta_1 a_s^2 -\beta_2 a_s^3 + {\cal O}\big(a_s^4\big)\,,
\eeq
where
\beeq\label{eq:betas}
\beta_0&=& \frac{11}{3}C_A-\frac{2}{3}N_f \;,\nn \\[2mm]
\beta_1&=& \frac{34}{3} C_A^2- \frac{10}{3} C_A N_f-2 C_F N_f \;,\nn\\[2mm]
\beta_2&=& \frac{2857}{54} C_A^3- \frac{1415}{54} C_A^2 N_f- \frac{205}{18} C_F C_A N_f+C_F^2 N_f+
\frac{79}{54} C_A N_f^2+\frac{11}{9} C_F N_f^2 \, ,
\eeeq
with $C_F=4/3$ and $C_A=3$. Keeping just the first term in each of Eqs.~(\ref{eq:split}) and~(\ref{eq:alpha})
yields the leading order (LO) evolution of the parton distributions. Taking into account also the second, or the second and 
third, terms corresponds to next-to-leading order (NLO) and next-to-next-to-leading order (NNLO) evolution, respectively.

The evolution equations may be simplified by introducing non-singlet and singlet combinations
of the quark and antiquark distributions; see e.g. Ref.~\cite{Furmanski:1981cw}. Following the notation of Ref.~\cite{Moch:2004pa} and
using charge conjugation invariance and flavor symmetry of QCD, we first write the evolution kernels 
$\Delta P_{ab}$ as 
\beeq
\Delta P_{q_i q_k}= \Delta P_{\bar{q}_i \bar{q}_k} &\equiv& \delta_{ik} \,\Delta P_{qq}^V + 
\Delta P_{qq}^S\;, \nn\\[2mm]
\Delta P_{q_i \bar{q}_k}= \Delta P_{\bar{q}_i q_k} &\equiv& \delta_{ik} \,\Delta P_{q\bar{q}}^V +  
\Delta P_{q\bar{q}}^S \;.
\eeeq
The splitting functions $\Delta P_{qq}^S$ and $\Delta P_{q\bar{q}}^S$ thus describe
splittings in which the flavor of the quark changes. Starting from
NNLO, $\Delta P_{qq}^S$ and $\Delta P_{q\bar{q}}^S$ differ \cite{Furmanski:1981cw,Catani:2004nc,Moch:2015usa}. 

We now introduce three types of flavor non-singlet combinations of parton densities:
\beq\label{eq:defnons}
\Delta q^{(V)} \,\equiv\, \sum_q \left( \Delta q - \Delta \bar{q} \,\right) \;\;,
\quad 
\Delta q^{(\pm)}\, \equiv\, \Delta q\pm \Delta \bar{q} - \frac{1}{N_f}
\sum_{q'} \left(\Delta  q' \pm \Delta \bar{q}{\,'} \right) \;,
\eeq
which turn out to diagonalize the evolution equations in the non-singlet sector. 
(Up to NLO it would be sufficient to consider only two non-singlet combinations;
owing to $\Delta P_{qq}^S\neq\Delta P_{q\bar{q}}^S$, it becomes necessary 
to introduce a third combination at NNLO and beyond.) Each of the three combinations
evolves in a simple closed form:
\beq\label{eq:evolnons}
\frac{d\,  \Delta  q^{(A)}(Q^2)}{d\ln Q^2}\, =\, \Delta P^{(A)}\big(\as(Q^2)\big) \; \Delta q^{(A)}(Q^2)
\;\;, \quad (A=V,\pm)  \;, 
\eeq
where the corresponding evolution kernels are
\beq\label{PVpm}
\Delta P^{(V)}\, =\, \Delta P_{qq}^V - \Delta P_{q\bar{q}}^V + N_f \left( \Delta P_{qq}^S - \Delta P_{q\bar{q}}^S
\right) \;\;,
\quad \Delta P^{(\pm)} \,=\, \Delta  P_{qq}^V \pm \Delta P_{q\bar{q}}^V \;.
\eeq
The decoupled non-singlet equations are trivial to solve; we will present the solutions
later. 

In the singlet sector defined by Eq.~(\ref{singsec}) we have coupled evolution equations
for $\Delta \Sigma$ and $\Delta G$: 
\begin{eqnarray}
\label{sieq}
\frac{d}{d\ln Q^2} \left( \begin{array}{c}
\Delta\Sigma (Q^2)\\[2mm]
\Delta G(Q^2) \end{array}\right) \,=\, 
\left( \begin{array}{cc}
\Delta P_{\Sigma\Sigma}\big(a_s(Q^2)\big) & 2 N_f \,\Delta P_{qG}\big(a_s(Q^2)\big) \\[2mm]
\Delta P_{Gq}\big(a_s(Q^2)\big) & \Delta P_{GG}\big(a_s(Q^2)\big) \end{array}\right)\,
\left( \begin{array}{c}
\Delta\Sigma (Q^2)\\[2mm]
\Delta G(Q^2) \end{array}\right)  \;,
\end{eqnarray}
where 
\beq\label{PSS}
\Delta P_{\Sigma\Sigma}\,\equiv\,\Delta P_{qq}^{V} + \Delta P_{q{\bar q}}^{V} + N_f
\big(\Delta P_{qq}^{S} + \Delta P_{q{\bar q}}^{S}\big)\,,
\eeq
and with the first moments of the splitting functions involving gluons, $\Delta P_{qG}$, $\Delta P_{Gq}$, 
$\Delta P_{GG}$. As we shall discuss below, thanks to the simplicity of the evolution kernels in the 
spin-dependent case the singlet equation may also be solved analytically in a simple way. 

The evolution of the helicity parton distributions is in itself closed and not affected by contributions from 
orbital angular momentum. On the other hand, $L_q$ and $L_g$ in Eq.~(\ref{JM}) are both scale
dependent and have their own evolution equations. As it turns out, their evolution is not closed but
is partly driven by $\Delta \Sigma(Q^2)$ and $\Delta G(Q^2)$. This has to be the case since the left-hand-side
of Eq.~(\ref{JM}) needs to remain independent of the scale. Presently, the evolution of $L_q$ and $L_g$ 
%WV new footnote, plus reference
is known only to lowest order~\cite{Ji:1995cu,Hagler:1998kg,Harindranath:1998ve}\footnote{We note that
the evolution for the total quark and gluon angular momenta in the Ji decomposition may be derived
by profiting from the relation between the total angular momentum operators and the quark and gluon energy 
momentum tensors~\cite{Ji:1996ek} and is actually known up to NNLO accuracy~\cite{Larin:1993vu,Larin:1996wd}.
Unfortunately, since there is no direct connection between the Ji and Jaffe-Manohar spin decompositions
(except for the quark spin piece), it is not possible to use these results to obtain the higher-order evolutions of
$L_q$ and $L_g$.}. Beyond LO, we therefore cannot separate the evolution of the two orbital 
components. However, we can still consider the evolution of the {\it total} orbital angular momentum
${\cal L}\equiv L_q+L_g$ by simply taking the derivative of Eq.~(\ref{JM}):
\beeq\label{OAMtot}
\frac{d \,{\cal L} (Q^2)}{d\ln Q^2}&=&-\frac{1}{2} \frac{d \,\Delta\Sigma (Q^2)}{d\ln Q^2}- \frac{d \,\Delta G(Q^2)}{d\ln Q^2}\nn\\[2mm]
&=&-\left(\frac{1}{2}\Delta P_{\Sigma\Sigma}+\Delta P_{Gq}\right)\Delta\Sigma (Q^2)-
\left(N_f \,\Delta P_{qG}+\Delta P_{GG}\right)\Delta G(Q^2) \,.
\eeeq
This relation will serve as an important cross-check for future calculations of the separate evolution
of $L_q$ and $L_g$ at higher orders. We note that, like for the helicity quark and antiquark distributions,
there will be a separate angular momentum piece $L_{q_i}$ for each flavor, and full evolution equations
will require introduction of non-singlet and singlet combinations. $L_q$ as appearing in the spin sum rule 
is the singlet. 

\section{First moments of the splitting functions}

We now collect the various $\Delta P_{ab}$ as available from the literature. At lowest order we have
\cite{Ahmed:1976ee,Altarelli:1977zs}
\beeq\label{LOP}
\Delta P^{(0)\pm}=\Delta P^{(0)S}_{qq} = \Delta P^{(0)S}_{q\bar{q}} &=& 0 \,,\nn \\[2mm]
\Delta P^{(0)}_{qG}&=&0 \,,\nn \\[2mm]
\Delta P^{(0)}_{Gq}&=&3 C_F \,,\nn \\[2mm]
\Delta P^{(0)}_{GG}&=& \beta_0 \,.
\eeeq
The second-order results in the $\overline{\mathrm{MS}}$ scheme 
may be found in Refs. \cite{Mertig:1995ny,Vogelsang:1995vh,Vogelsang:1996im}.
\beeq\label{NLOP}
\Delta P^{(1)+}&=&  0 \,, \nn \\[2mm]
\Delta P^{(1)-}&=& C_F (C_A-2 C_F) \left(-13+12 \zeta_2-8 \zeta_3\right) \,, \nn \\[2mm]
N_f\left( \Delta P^{(1)S}_{qq} + \Delta P^{(1)S}_{q\bar{q}}\,\right) &=& - 2 N_f \,  \Delta P^{(0)}_{Gq} \,, \nn \\[2mm]
N_f\left(\Delta P^{(1)S}_{qq}  - \Delta P^{(1)S}_{q\bar{q}}\right) &=&  0\,,\nn \\[2mm]
\Delta P^{(1)}_{qG}&=&0\,, \nn \\[2mm]
\Delta P^{(1)}_{Gq}&=&\frac{71}{3} C_F C_A -9 C_F^2-\frac{2}{3} C_F N_f \,,\nn \\[2mm]
\Delta P^{(1)}_{GG}&=& \beta_1\,,
\eeeq
with $\zeta_i\equiv\zeta(i)$ the respective value of Riemann's zeta function. Finally, at 
NNLO we have from Refs. \cite{Kodaira:1979pa,Moch:2014sna, Moch:2015usa}:
\beeq
%\label{NNLOP}
\Delta P^{(2)+}&=& 0 \,,\nn \\[2mm]
\Delta P^{(2)-}&=&  \Bigg\{ C_F \left(\frac{145}{2}-62 \zeta_2+164 \zeta_3-372 \zeta_4+
                               48 \zeta_2 \zeta_3+208 \zeta_5 \right) \nn \\[2mm]
               &+& C_A\left(\frac{1081}{36} +\frac{245}{3} \zeta_2-\frac{3214}{9} \zeta_3 +
                                \frac{1058}{3} \zeta_4 -48 \zeta_2 \zeta_3 -112 \zeta_5\right)\nn \\[2mm]
               &-& N_f  \left(\frac{76}{9}+\frac{44}{3} \zeta_2 -\frac{448}{9} \zeta_3 +\frac{68}{3} \zeta_4\right) 
               \Bigg\}\,C_F(C_A-2 C_F)\,,  \nn %\\[2mm]
\eeeq
\beeq\label{NNLOP}               
N_f\left( \Delta P^{(2)S}_{qq} + \Delta P^{(2)S}_{q\bar{q}}\right) &=& - 2 N_f \,  \Delta P^{(1)}_{Gq}\,,\nn \\[2mm]
N_f\left(\Delta P^{(2)S}_{qq}  - \Delta P^{(2)S}_{q\bar{q}}\right) &=&  
\frac{8 N_f}{C_A}\, d_{abc} d_{abc} \,\big(23-12 \zeta_2-16 \zeta_3\big) \,,\nn \\[2mm]
\Delta P^{(2)}_{qG}&=&0 \,, \nn \\[2mm]
\Delta P^{(2)}_{Gq}&=& \frac{1607}{12} C_F C_A^2 - \frac{461}{4} C_F^2 C_A + \frac{63}{2} C_F^3 \nn \\[2mm]
               &+& \left(\frac{41}{3}-72 \zeta_3\right) C_F C_A N_f
       -\left( \frac{107}{2}-72 \zeta_3\right) C_F^2 N_f-\frac{13}{3} C_F N_f^2\,, \nn \\[2mm]
\Delta P^{(2)}_{GG}&=& \beta_2 \,.
\eeeq
In the above equations, $C_F=4/3$, $C_A=3$, $N_f$ is the number of flavors, and 
$d_{abc} d_{abc}/C_A=5/18$. 

There are systematic patterns among the above results which may be understood from 
general arguments. First of all, $\Delta P^+$ has to vanish to all orders in the strong coupling.
As follows from Eqs.~(\ref{eq:defnons},\ref{eq:evolnons}), $\Delta P^+$ governs the
evolution of combinations such as $\Delta u +\Delta \bar{u} - (\Delta d +\Delta \bar{d})$,
which correspond to matrix elements of flavor non-singlet axial currents. Such matrix elements 
%WV small changes:
are not renormalized and are hence scale independent~\cite{Kodaira:1978sh} to all 
orders, consistent with the Bjorken sum rule~\cite{Bjorken:1966jh}.\footnote{Evidently, in a perturbative calculation, 
one could in principle choose a factorization scheme in which $\Delta P^+$ becomes non-zero at NLO or beyond.
However, such a scheme would be unphysical~\cite{Stratmann:1995fn}.}

As is well known (see Eqs.~(\ref{sieq},\ref{PSS})), the vanishing of $\Delta P^+$ immediately implies 
that the evolution of the singlet $\Delta \Sigma$ is to all orders driven by the ``pure-singlet'' anomalous
dimension $N_f\big(\Delta P_{qq}^{S} + \Delta P_{q{\bar q}}^{S}\big)$. The explicit results shown in 
the above equations suggest that 
\beq
N_f\left( \Delta P^{(j+1)S}_{qq} + \Delta P^{(j+1)S}_{q\bar{q}}\right) \,=\, - 2 N_f \,  \Delta P^{(j)}_{Gq}\,,
\eeq
or, generalized to all orders,
\beq\label{gen1}
\Delta P_{\Sigma\Sigma} \,=\, - 2 N_f \, a_s\,\Delta P_{Gq}\,.
\eeq
Furthermore, we deduce from Eqs.~(\ref{NLOP},\ref{NNLOP})
\beeq\label{gen2}
\Delta P_{qG}&=&0\,,\nn\\[2mm]
\Delta P_{GG}&=&-\frac{\beta(a_s)}{a_s}\,.
\eeeq
The all-order results just given may in fact be understood by an argument
given in Ref.~\cite{Altarelli:1990jp}. The quark singlet combination $\Delta \Sigma$ 
corresponds to the proton matrix element of the flavor-singlet axial current, 
\beq
S^\mu\,\Delta \Sigma\,=\,\bra{P,S\,}\,\bar{\psi} \,\gamma^\mu \gamma_5 \,\psi\,\ket{\,P,S}
\,\equiv\,\bra{P,S\,}\,j_5^\mu \,\ket{\,P,S}\,,
\eeq
where $S$ is the proton's polarization vector. Because of the axial anomaly, the singlet axial current is not 
conserved:
\beeq\label{anomaly}
\partial_\mu \,j_5^\mu&=&2N_f\,a_s\,{\mathrm{Tr}}\Big[F_{\mu\nu}\,\tilde{F}^{\mu\nu}\Big]\nn\\[2mm]
&=&2N_f\,a_s\,\partial_\mu\left\{\varepsilon^{\mu\nu\rho\sigma}\,{\mathrm{Tr}}\left[A^\nu\left(
F^{\rho\sigma}-\frac{2}{3}\,A^\rho\,A^\sigma\right)\right]\right\}\nn\\[2mm]
&\equiv&2N_f\,a_s\,\partial_\mu K^\mu\,.
\eeeq
In the first line, $F^{\mu\nu}$ is the gluonic field strength tensor and $\tilde{F}^{\mu\nu}$ its dual.
In the second line we have used that ${\mathrm{Tr}}[F\tilde{F}\,]$ may be written as the divergence 
of the ``anomalous current'' that we denote by $K$. From Eq.~(\ref{anomaly}) we conclude that
$j_5^\mu-2N_f\,a_s K^\mu$ is conserved:
\beq\label{JK}
\partial_\mu\Big(j_5^\mu-2N_f\,a_s K^\mu\Big)\,=\,0\,.
\eeq
The relation $\partial_\mu \,j_5^\mu=2N_f\,a_s\,{\mathrm{Tr}}[F\tilde{F}\,]$ holds to all orders in perturbation theory.
As a result, Eq.~(\ref{JK}) holds to all orders as well. As was discussed in~\cite{Altarelli:1990jp}, in perturbation
theory we may relate matrix elements of $K^\mu$ to the gluon spin contribution:
$S^\mu\,\Delta G=-\bra{P,S\,}\,K^\mu \,\ket{\,P,S}$. Although $K$ depends on the choice of gauge,
its forward proton matrix element is gauge invariant, except for topologically nontrivial gauge 
transformations that change the winding number. (The latter feature makes the identification
of $\Delta G$ with the matrix element of $K$ impossible beyond perturbation theory~\cite{Jaffe:1989jz}.)
From the conservation law in~(\ref{JK}) we may thus conclude
\beq
\frac{d}{d\ln Q^2} \left( \Delta\Sigma (Q^2)+2N_f\,a_s(Q^2)\,\Delta G(Q^2)\right)\,=\,0\,.
\eeq
Inserting the general evolution equations for $\Delta \Sigma$ and $\Delta G$ in~(\ref{sieq}),
as well as the renormalization group equation for $a_s(Q^2)$ in~(\ref{eq:alpha}),
we find, on the other hand,
\beeq\label{eq:vanish}
\frac{d}{d\ln Q^2} \left( \Delta\Sigma+2N_f\,a_s\,\Delta G\right)&=&
\Big(\Delta P_{\Sigma\Sigma}+2 N_f a_s \,\Delta P_{Gq}\Big)\Delta \Sigma\nn\\[2mm]
&+&\left(2 N_f\Delta P_{qG}+\Delta P_{GG}+\frac{\beta(a_s)}{a_s}\right)2N_fa_s\Delta G\,.
\eeeq
The right-hand-side vanishes when the all-order relations given in Eqs.~(\ref{gen1}) and~(\ref{gen2})
are satisfied. Equivalent results are found when studying the renormalization of the axial anomaly 
in dimensional regularization in \cite{Larin:1993tq}. One may object that Eqs.~(\ref{gen2})
do not follow from~(\ref{eq:vanish}) in a strict mathematical sense; however, there is little (if any) freedom
physically to obtain results other than~(\ref{gen2}) from the last term in~(\ref{eq:vanish}). In particular, 
the $C_A$ parts in $P_{GG}$ can only be canceled by those in the $\beta$-function. The 
explicit verification to three loops by the results of~\cite{Moch:2014sna,Moch:2015usa} is
of course a strong argument for the all-order validity of~(\ref{gen2}). In addition, the 
vanishing of $\Delta P_{qG}$ in any physical scheme is a consequence of helicity conservation.

We note that as seen in Refs.~\cite{Moch:2014sna,Moch:2015usa,Stratmann:1995fn} relations like 
$\Delta P^+=0$ and~(\ref{gen1}) and~(\ref{gen2}) may not emerge automatically in an actual 
higher-loop calculation of the splitting functions, where dimensional regularization and a prescription for 
$\gamma_5$ and the Levi-Civit\`{a} tensor have to be adopted. They may then be reinstated by a 
factorization scheme transformation, so that the correct physical splitting functions are obtained. 

It is now clear that in the $\overline{\rm MS}$ scheme a {\it single} anomalous dimension, 
$\Delta P_{\Sigma\Sigma}$, resulting from the axial anomaly, governs the evolution of the 
quark and gluon spin contributions and (via Eq.~(\ref{OAMtot})) of the total orbital angular momentum.
Inserting our findings into Eq.~(\ref{sieq}), we obtain
\beq
\label{sieq1}
\frac{d}{d\ln Q^2} \left( \begin{array}{c}
\Delta\Sigma\\[2mm]
\Delta G \end{array}\right) \,=\, 
\left( \begin{array}{cc}
\Delta P_{\Sigma\Sigma}(a_s)& 0 \\[2mm]
-\frac{1}{2N_f a_s} \, \Delta P_{\Sigma\Sigma}(a_s)&-\frac{\beta(a_s)}{a_s} \end{array}\right)\,
\left( \begin{array}{c}
\Delta\Sigma \\[2mm]
\Delta G \end{array}\right)  \;,
\eeq
where we have dropped the ubiquitous argument $Q^2$. We may further simplify this equation by
defining~\cite{Altarelli:1990jp}
\beq
\Delta \Gamma(Q^2)\,\equiv\,a_s(Q^2)\Delta G(Q^2)\,.
\eeq
From~(\ref{sieq1}) we then have
\beq
\label{sieq2}
\frac{d}{d\ln Q^2} \left( \begin{array}{c}
\Delta\Sigma\\[2mm]
\Delta \Gamma \end{array}\right) \,=\, 
\left( \begin{array}{cc}
\Delta P_{\Sigma\Sigma}(a_s)& 0 \\[2mm]
-\frac{1}{2N_f} \, \Delta P_{\Sigma\Sigma}(a_s)& 0 \end{array}\right)\,
\left( \begin{array}{c}
\Delta\Sigma \\[2mm]
\Delta \Gamma \end{array}\right)  \;.
\eeq
The lower right entry of the evolution matrix now vanishes
since in the product $a_s(Q^2)\Delta G(Q^2)$ the evolution of the strong coupling exactly 
cancels the $\Delta P_{GG}$ part of the evolution of $\Delta G$. Clearly, Eq.~(\ref{sieq2}) is straightforward
to solve, and we will return to the equation shortly. 

Thanks to Eq.~(\ref{gen1}) we may now determine the {\it four-loop} (N$^3$LO) contribution to the
anomalous dimension $\Delta P_{\Sigma\Sigma}$ from the three-loop value 
$\Delta P_{Gq}^{(2)}$ computed in Ref. \cite{Moch:2015usa}:
\beeq
\Delta P_{\Sigma\Sigma}^{(3)}\;=\;- 2 N_f \,  \Delta P^{(2)}_{Gq}&=&
 - 2 N_f C_F \left[\,\frac{1607}{12} C_A^2 - \frac{461}{4} C_F C_A + \frac{63}{2} C_F^2\right. \nn \\[2mm]
               &+&\left.  \left(\frac{41}{3}-72 \zeta_3\right) C_A N_f
       -\left( \frac{107}{2}-72 \zeta_3\right) C_F N_f-\frac{13}{3} N_f^2\,\right].
\eeeq

\section{Higher-order solutions in the singlet sector}

We now proceed to solve the singlet evolution equation~(\ref{sieq2}). 
Changing $d \ln Q^2$ to $d a_s$ via Eq.~(\ref{eq:alpha}), we have
for the evolution of $\Delta\Sigma$, up to N$^3$LO:
\beeq
\label{eq:singeq}
\frac{d\ln\Delta\Sigma (Q^2)}{da_s(Q^2)}&=&-\frac{ \Delta P_{\Sigma\Sigma}^{(0)} +a_s  \, \Delta P_{\Sigma\Sigma}^{(1)} 
+a_s^2\, \Delta P_{\Sigma\Sigma}^{(2)}+a_s^3\, \Delta P_{\Sigma\Sigma}^{(3)}}{a_s \beta_0 +a_s^2 \,\beta_1 +a_s^3\, \beta_2}
\nn\\[2mm]
&=&-\frac{\Delta P_{\Sigma\Sigma}^{(1)} 
+a_s\, \Delta P_{\Sigma\Sigma}^{(2)}+a_s^2\, \Delta P_{\Sigma\Sigma}^{(3)}}{\beta_0 +a_s \,\beta_1 +a_s^2\, \beta_2}\,,
\eeeq
where in the second line we have used that $\Delta P_{\Sigma\Sigma}^{(0)}=0$, since the evolution of $\Delta\Sigma$
starts only at NLO. Expanding the right-hand side of Eq.~(\ref{eq:singeq}) up to second order it becomes
\beeq
\label{eq:singexp}
\frac{d\ln \Sigma (Q^2)}{da_s(Q^2)} = \Bigg[ \! &-&\!\frac{\Delta P_{\Sigma\Sigma}^{(1)}}{\beta_0}  \,+\,
\frac{a_s}{\beta_0^2} \left( \beta_1 \,\Delta P_{\Sigma\Sigma}^{(1)}  -\beta_0 \,\Delta P_{\Sigma\Sigma}^{(2)} \right) \nn \\[2mm]
 \!&+&\! \frac{a_s^2}{\beta_0^3}  \Bigg(  -\beta_1^2 \, \Delta P_{\Sigma\Sigma}^{(1)} \,+\,
 \beta_0 \beta_2 \,\Delta P_{\Sigma\Sigma}^{(1)}+\beta_0 \beta_1 \,\Delta P_{\Sigma\Sigma}^{(2)}- \beta_0^2 \,
 \Delta P_{\Sigma\Sigma}^{(3)} \Bigg)
  \Bigg]\,.
\eeeq
This equation is readily solved analytically. The solution gives the first moment of the singlet at scale $Q$ in terms of
its boundary value at the ``input'' scale $Q_0$:
 \beeq
\label{eq:singsol}
\frac{ \Delta\Sigma (Q^2)}{ \Delta \Sigma (Q_0^2)} &=& \exp{[0]} \;\times\;
\exp
 \Bigg[  - \frac{ a_Q -a_{0}}{\beta_0} \,\Delta P_{\Sigma\Sigma}^{(1)} \Bigg] \;\times\;
 \exp
\Bigg[   \frac{ a_Q^2 -a_{0}^2}{2\beta_0^2} \,  \left( \beta_1 \,\Delta P_{\Sigma\Sigma}^{(1)}  -\beta_0\,
\Delta P_{\Sigma\Sigma}^{(2)}\right) \Bigg] \nn\\[2mm]
&&\!\times\;\exp
\Bigg[   \frac{ a_Q^3 -a_{0}^3}{3\beta_0^3} \left(
- \beta_1^2 \, \Delta P_{\Sigma\Sigma}^{(1)}+\beta_0 \beta_2 \, \Delta P_{\Sigma\Sigma}^{(1)}+\beta_0\beta_1 \,
\Delta P_{\Sigma\Sigma}^{(2)}- \beta_0^2 \,\Delta P_{\Sigma\Sigma}^{(3)}\right)\Bigg]
\nn\\[2mm]
 &\equiv& K^{{\mathrm{(LO)}}}\; \times\;  K^{{\mathrm{(NLO)}}}\;\times\;K^{{\mathrm{(NNLO)}}} \;\times\; K^{{\mathrm{(N}}^3{\mathrm{LO)}}}\,,
\eeeq
where $a_Q\equiv a_s(Q^2)$ and $a_0\equiv a_s(Q_0^2)$. For completeness, we have included the LO term, 
which predicts a constant $\Delta \Sigma$. 

%%%%%%%%%%%%%%%%%%%%%%%%%%%%%%%%%%%%%%%%%%%%%%%%%%%
\begin{figure}[h!]
\vspace*{4mm}
 \centerline{  \epsfig{figure=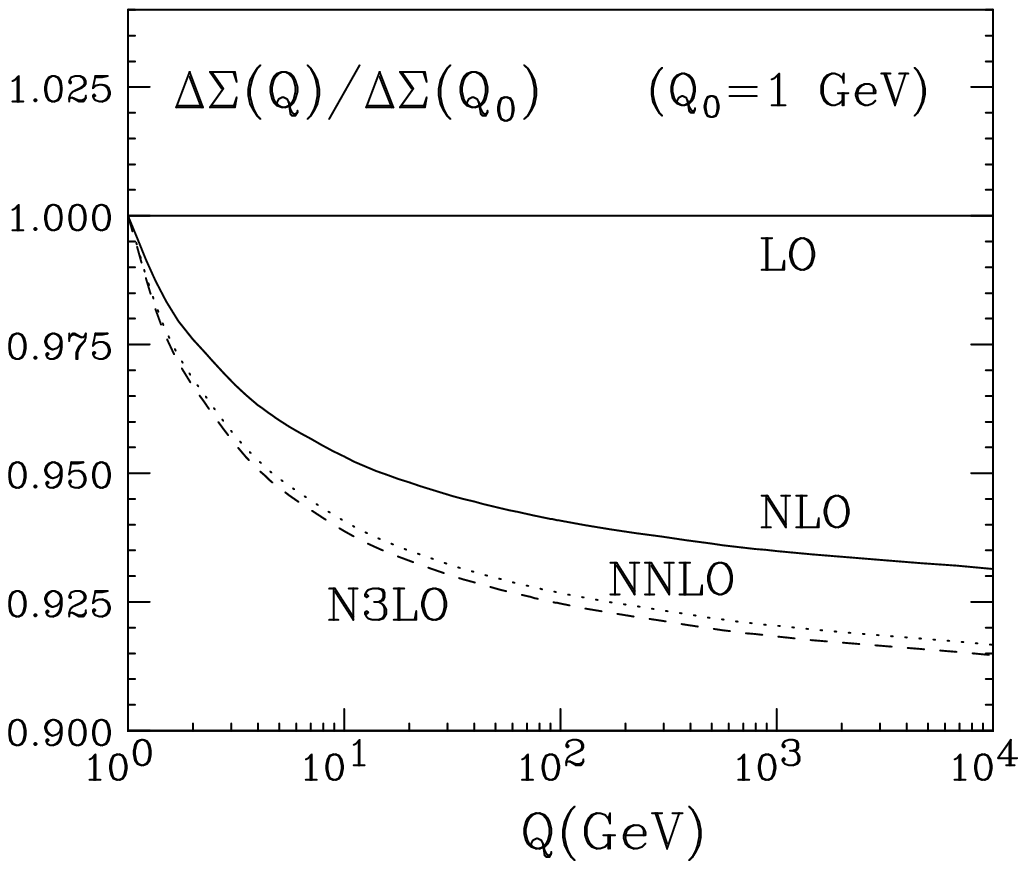,width=0.55\textwidth,clip=}}
\caption{ \label{singlet} {\it Evolution of the first moment of the polarized singlet distributions at LO, NLO, 
NNLO and N$^3$LO, starting from the initial scale $Q=1$ GeV.}}
\end{figure}                 
%%%%%%%%%%%%%%%%%%%%%%%%%%%%%%%%%%%%%%%%%%%%%%%%%%%

Figure \ref{singlet} shows the quark singlet evolution factor on the right-hand-side of Eq. (\ref{eq:singsol}), 
assuming a fixed number $N_f=3$ in the anomalous dimensions and the beta function,
%WV inserted footnote:
and using the full NNLO evolution of the coupling constant\footnote{Alternatively, one could use
at each order a coupling constant defined by truncating the QCD $\beta$-function to that order.
This approach would mostly affect the LO results, since the coupling constant at this order is larger. 
It would also slightly limit the range of applicability of the LO calculation in the backward evolution 
since the non-perturbative regime would be reached already at higher scales. Nevertheless, the main 
results of this paper would remain basically unchanged since the values of the coupling constant 
are quite similar at NLO, NNLO and beyond.}. We choose a relatively low input scale  $Q_0=1$,
with a value $\alpha_s(Q_0)=0.404$ \footnote{That value corresponds to the conventional $\alpha_s(M_Z)=0.1181$ for $N_f=5$.
In this paper we always use the NNLO expression for the coupling constant, independently of the order considered,
as a way of isolating the effect of the higher order splitting functions in the corresponding evolution.}.
One can see that the NLO evolution affects the quark spin content of the proton by up to $7\%$
while NNLO evolution adds an extra $\sim 1-2\%$ effect. The numerical impact of the four-loop
term $\Delta P_{\Sigma\Sigma}^{(3)}$ reaches only ${\cal{O}}(0.2\%)$ at the highest scale.

\begin{figure}[t!]
\vspace*{4mm}
 \centerline{  \epsfig{figure=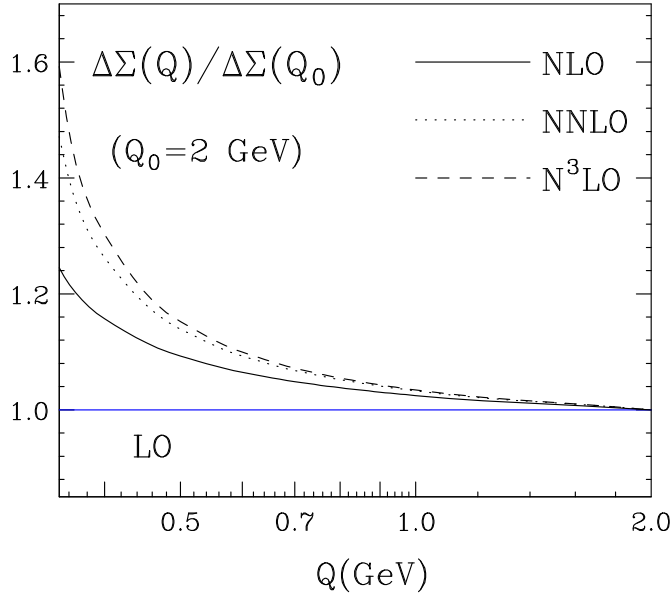,width=0.55\textwidth,clip=}}
\caption{ \label{singletback} {\it Backward evolution of the first moment of the polarized singlet distributions at LO, NLO, NNLO and N$^3$LO, starting from the initial scale $Q=2$ GeV.}}
\end{figure}

Ultimately, as discussed in Ref.~\cite{Altenbuchinger:2010sz,Jaffe:1987sx}, one may want to compare helicity
parton distribution functions extracted from experiment or computed on the lattice~\cite{Lin:2017snn} with 
calculations performed in QCD-inspired models of nucleon structure. The latter typically 
are formulated at rather low momentum scales of order of a few hundred MeV. Given the high
order of perturbation theory now available for evolution, it is therefore interesting to evolve
the singlet spin contributions not only to large perturbative scales, but also ``backward'' 
towards the limit of validity of perturbation theory~\cite{Altenbuchinger:2010sz}. 
In Fig.~\ref{singletback} we show the evolution of $\Delta\Sigma$ at LO, NLO, NNLO and N$^3$LO
down to $Q\sim 0.35$ GeV, starting from the initial scale $Q=2$ GeV. 
Since at low scales the  approximate analytical expressions for the running of the coupling constant 
$\alpha_s$ start to deviate from the exact result, we rely on the accurate numerical solution of 
Eq.~(\ref{eq:alpha}) to NNLO accuracy. As can be observed, and as is expected, the higher order 
terms affect the evolution of the singlet in a significant way, much more strongly than what we 
found for the evolution to larger scales. On the other hand, a striking feature is that the 
evolution remains relatively stable even down to such low scales as considered here: 
At the lower end of Figure \ref{singletback} the N$^3$LO contribution enhances the singlet by 
a modest 8\% compared to the previously known NNLO result, despite the fact that 
at $Q=0.35$ GeV the coupling constant becomes $\alpha_s\sim 1.3$, precariously
past the boundaries of the perturbative domain. In addition, all higher orders
(NLO, NNLO, N$^3$LO) go in the same direction. We note that the upturn of $\Delta\Sigma$
toward small scales -- in the direction of large quark and anti-quark spin contributions to the proton spin --
was already observed to NLO and NNLO in Refs.~\cite{Jaffe:1987sx} and~\cite{Altenbuchinger:2010sz},
respectively. We also remark that results on high-loop evolution may be useful for lattice-QCD
studies of nucleon structure, possibly allowing cross-checks of the nonperturbative renormalization
carried out on the lattice.  

%%%%%%%%%%%%%%%%%%%%%%%%%%%%%%%%%%%%%%%%%%%%%%%%%%%
%%%%%%%%%%%%%%%%%%%%%%%%%%%%%%%%%%%%%%%%%%%%%%%%%%%

The solution of the evolution equation for the gluon spin contribution now follows directly. 
From the lower row in Eq.~(\ref{sieq2}) we have by simple integration and using
again $d\ln Q^2=da_s/\beta(a_s)$
\beeq\label{DGamsol}
\Delta \Gamma(Q^2) &=&\Delta\Gamma(Q_0^2)-\int_{a_0}^{a_Q} da_s\,
\frac{\Delta P_{\Sigma\Sigma}(a_s)}{2N_f \beta(a_s)}\,\Delta\Sigma(Q^2)\nn\\[2mm]
&=&\Delta\Gamma(Q_0^2)+\int_{a_0}^{a_Q} \frac{da_s}{\beta(a_s)}\,a_s\,
\Delta P_{Gq}(a_s)\,\Delta\Sigma(Q^2)\,,
\eeeq
where again $a_Q\equiv a_s(Q^2)$ and $a_0\equiv a_s(Q_0^2)$, and
where in the second line we have used Eq.~(\ref{gen1}) to replace $\Delta P_{\Sigma\Sigma}$ by $\Delta P_{Gq}$,
which is more natural in the case of the gluon distribution. 

An immediate observation is that the integral on the right-hand-side of~(\ref{DGamsol}) starts at order
$a_s(Q^2)$ and $a_s(Q_0^2)$. Therefore, we arrive at the well-known 
result~\cite{Ratcliffe:1987dp,Ramsey:1988dm,Altarelli:1988nr} 
that the leading term in $\Delta\Gamma$ is a constant in $Q^2$, so that the 
first moment of the gluon spin contribution evolves as the inverse of the strong coupling.
Inserting the solution for $\Delta\Sigma(Q^2)$ from Eq.~(\ref{eq:singsol}) into~(\ref{DGamsol}) and
carrying out the integration, we find the full NNLO analytical solution
\beq\label{dGevol}
\Delta G(Q^2)\,=\,\frac{a_s(Q_0^2)}{a_s(Q^2)}\,\Delta G(Q_0^2)\,+\,\Delta\Sigma(Q_0^2)\,F\Big(a_s(Q^2),a_s(Q_0^2)\Big)\,,
\eeq
where
\beq
F(a_Q,a_0)\,=\,F^{{\mathrm{LO}}}\left(\frac{a_0}{a_Q}\right)\,+\,
a_Q\,F^{{\mathrm{NLO}}}\left(\frac{a_0}{a_Q}\right)\,+\,
a_Q^2\,F^{{\mathrm{NNLO}}}\left(\frac{a_0}{a_Q}\right)\,,
\eeq
with
\beeq
\label{solgluon}
F^{{\mathrm{LO}}}(r)&=& -(1-r)\frac{\Delta P^{(0)}_{Gq}}{\beta_0} \,,\nn \\[2mm]
F^{{\mathrm{NLO}}}(r)&=& \frac{1-r^2}{2\beta_0^2} \left(\beta_1 \,\Delta P^{(0)}_{Gq}-\beta_0\, \Delta P^{(1)}_{Gq}\right)
+\frac{(1-r)^2}{2\beta_0^2}\,\Delta P^{(0)}_{Gq} \,\Delta P_{\Sigma\Sigma}^{(1)} \,,\nn \\[2mm]
F^{{\mathrm{NNLO}}}(r)&=& \frac{1-r^3}{3\beta_0^3} 
\left(\beta_0 \beta_2 \,\Delta P^{(0)}_{Gq} - \beta_1^2 \,\Delta P^{(0)}_{Gq}  + \beta_0 \beta_1 \,\Delta P^{(1)}_{Gq}   - 
\beta_0^2\,\Delta P^{(2)}_{Gq}          \right) \nn \\[2mm]
&+& \frac{(1-r)^2} {6\beta_0^3} \left[  -\,3 (1+r) \beta_1 \,\Delta P^{(0)}_{Gq}  \,\Delta P_{\Sigma\Sigma}^{(1)} + 
(2+r) \beta_0 \,\Delta P^{(1)}_{Gq}  \,\Delta P_{\Sigma\Sigma}^{(1)}\right.\nn \\[2mm]
&& \left.\hspace{1.68cm} +\, (1+2r)\,\beta_0 \,\Delta P^{(0)}_{Gq} \,\Delta P_{\Sigma\Sigma}^{(2)} 
\right]\,-\,
\frac{(1-r)^3} {6\beta_0^3}\,\Delta P^{(0)}_{Gq}\,\big(\Delta P_{\Sigma\Sigma}^{(1)}\big)^2\,.
\eeeq
We note that in contrast to $\Delta\Sigma$, we can only give the NNLO evolution of $\Delta G$ here.
This is due to the fact that $\Delta \Gamma$ is shifted by one power of $a_s$ relative to $\Delta G$.
In order to obtain the N$^3$LO solution for $\Delta G$ one would need the four-loop splitting kernel 
$\Delta P^{(3)}_{Gq}$ which is presently still unavailable. 

%%%%%%%%%%%%%%%%%%%%%%%%%%%%%%%%%%%%%%%%%%%%%%%%%%%
\begin{figure}[h]
\vspace*{4mm}
 \centerline{  \epsfig{figure=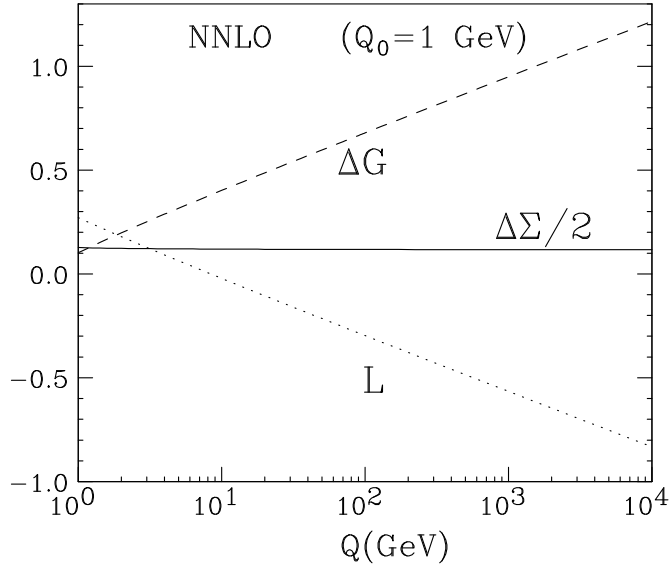,width=0.55\textwidth,clip=}}
\caption{ \label{fig:all} {\it Evolution of the quark and gluon spin contributions $\frac{1}{2}\Delta\Sigma$
and $\Delta G$ at NNLO, starting from the inital scale $Q=1$ GeV. We also show the evolution of
${\cal L}$ following from Eq.~(\ref{OAMtot}).}}
\end{figure}                                            
%%%%%%%%%%%%%%%%%%%%%%%%%%%%%%%%%%%%%%%%%%%%%%%%%%%    

Figure \ref{fig:all} shows the NNLO evolution of the gluon spin contribution to the proton spin, starting from 
the values $\Delta G=0.102$ and $\Delta\Sigma=0.254$ at $Q_0=1$ as realized in the 
global analysis~\cite{deFlorian:2014yva}. We also show the evolution of $\frac{1}{2}\Delta\Sigma$
and the evolution of the total orbital angular momentum ${\cal L}$ resulting from~(\ref{OAMtot}). 
Notice that both $\Delta G$ and ${\cal L}$ have a divergent behaviour at large scales, 
resulting in a rather unphysical cancellation of two very large contributions to fulfill the spin sum rule.

%%%%%%%%%%%%%%%%%%%%%%%%%%%%%%%%%%%%%%%%%%%%%%%%%%%
\begin{figure}[h]
\vspace*{4mm}
 \centerline{  \epsfig{figure=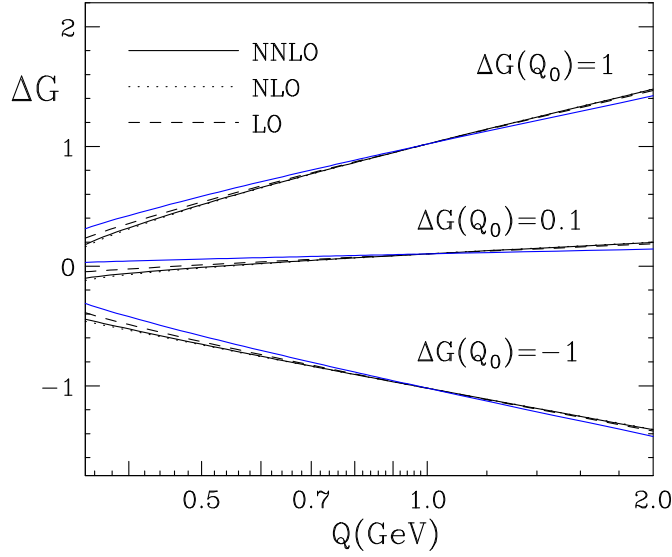,width=0.55\textwidth,clip=}}
\caption{ \label{fig:gback} {\it Backward evolution of the gluon spin contribution $\Delta G$ at LO (dashes), NLO (dots) and NNLO (solid line), 
starting from three different scenarios at the inital scale $Q_0=1$ GeV: $\Delta G(Q_0)=+1,0.1,-1$. The blue solid line corresponds to the leading $1/a_s$ term in Eq.~(\ref{dGevol})}.}
\end{figure}                                            
%%%%%%%%%%%%%%%%%%%%%%%%%%%%%%%%%%%%%%%%%%%%%%%%%%%       
 
 As we discussed above for the singlet contribution, it is also interesting to analyze the behavior of the gluonic spin contribution
 at lower scales. In Fig. \ref{fig:gback} we show the backward evolution of $\Delta G$ at LO (dashes), NLO (dots) and NNLO (solid line) for three different
 scenarios, corresponding to  setting $\Delta G(Q_0)=+1,0.1,-1$ at the initial scale  $Q_0=1$ GeV \footnote{$\Delta G(Q_0)=0.1$ corresponds to the result of the DSSV fit in \cite{deFlorian:2014yva}.}. For each scenario, we observe a striking convergence of the fixed order results down to very low scales, always towards small gluonic contributions. Even though the  ''$F$'' term in Eq.~(\ref{dGevol}) contains corrections proportional to positive powers of $\alpha_s$  that could spoil the convergence of the expansion in the non-perturbative region, the evolution of the gluonic contribution is completely dominated by the leading $1/a_s$ term in Eq.~(\ref{dGevol}), as can be observed in 
  Figure \ref{fig:gback} where we also present this term separately 
   for each scenario.  Our findings set a strong constraint on the proton spin content carried by gluons at {\it hadronic} scales.
Within the rather extreme scenarios analyzed here (for which the gluon contribution accounts for 
as much as twice the spin of the proton at  $Q_0=1$ GeV!), we obtain the requirement $ |\Delta G(Q\sim0.35 \,{\rm GeV})| \lesssim 0.3$. 
Indeed, the few available model estimates of $\Delta G$ suggest values of the order $0.2-0.3$~\cite{Jaffe:1995an,Barone:1998dx,Lee:2000mv,Chen:2006ng,Bass:2011zn} at a low hadronic scale.

 \section{``Static'' value of $\Delta G$}
 
As we have discussed, $\Delta G(Q^2)$ in general evolves as $1/a_s(Q^2)$ for large scales. 
As inspection of Eq.~(\ref{dGevol}) shows, depending on the input values of $\Delta G(Q_0^2)$ 
and $\Delta \Sigma(Q_0^2)$ the evolution can be towards large positive or negative values. 
This implies that there is a specific input, a ``critical point'', for which $\Delta G(Q^2)$ actually remains 
almost constant~\cite{deFlorian:2009vb} and tends to a finite asymptotic value as $Q^2\to\infty$.
This ``static'' value of $\Delta G$ is expected to change 
from order to order in perturbation theory. To determine it at a given order, we only need to tune the input 
such that the $1/a_s(Q^2)$ term in the solution for $\Delta G(Q^2)$ is canceled. Starting
from Eq.~(\ref{dGevol}) we demand
\beq
a_s(Q_0^2)\,\Delta G(Q_0^2)\,+\,a_s(Q^2)\,\Delta\Sigma(Q_0^2)\,F\Big(a_s(Q^2),a_s(Q_0^2)\Big)
\,=\, {\cal O}\big(a_s(Q^2)\big)\,.
\eeq
To LO, using~(\ref{solgluon}), this condition becomes
\beq
\Delta G_{\mathrm{stat}}^{{\mathrm{LO}}}(Q_0^2) \,=\, -  \Delta\Sigma(Q_0^2) \,  
\frac{\Delta P^{(0)}_{Gq}}{\beta_0} =-\frac{4}{9} \,\Delta\Sigma(Q_0^2)  \,\simeq \,-0.113\,,
\eeq
where we have used $N_f=3$ flavors and again $\Delta\Sigma(Q_0^2=1\,{\mathrm{GeV}}^2) = 0.254$. 
The gluon spin contribution then remains constant at the value $\Delta G_{\mathrm{stat}}^{{\mathrm{LO}}}(Q_0^2)$.

At NLO, the necessary input value for the static solution becomes
\beeq
\label{staticnlo}
\Delta G_{\mathrm{stat}}^{{\mathrm{NLO}}}(Q_0^2) &=& -  \Delta\Sigma(Q_0^2)  \left[ \, 
\frac{\Delta P^{(0)}_{Gq}}{\beta_0} 
+a_0  \frac{-\beta_1 \,\Delta P^{(0)}_{Gq} +\beta_0\, \Delta P^{(1)}_{Gq}  +\Delta P^{(0)}_{Gq} \,
\Delta P_{\Sigma\Sigma}^{(1)}}{2\beta_0^2}\,\right] \nn \\[2mm]
&=& - \left[\, \frac{4}{9} + \frac{166}{81} \,a_0\,\right] \Delta\Sigma(Q_0^2)\,  \simeq\, -0.13\,.
\eeeq
The NLO ``static'' solution is no longer completely constant in $Q^2$. However, by construction
it does converge asymptotically to a finite value, given by
\beeq
 \label{staticnloinf}
\Delta G_{\mathrm{stat}}^{{\mathrm{NLO}}}(\infty) &=& -  \Delta\Sigma(Q_0^2)  \left[  \,\frac{\Delta P^{(0)}_{Gq}}{\beta_0} 
+a_0 \, \frac{ \Delta P^{(0)}_{Gq} \,\Delta P_{\Sigma\Sigma}^{(1)}}{\beta_0^2}
\right] \nn \\[2mm]
&=& - \left[\, \frac{4}{9} - \frac{32}{27}\, a_0\,\right] \Delta\Sigma(Q_0^2) \, \simeq \,-0.103\,.
\eeeq
We note that a value of similar size was in fact found in the early NLO DSSV 
analysis~\cite{wvtalk,deFlorian:2009vb,deFlorian:2008mr}. 

Finally, at NNLO, the corresponding values are
\beeq
 \label{staticnlo1}
\Delta G_{\mathrm{stat}}^{{\mathrm{NNLO}}}(Q_0^2) &=& -  \Delta\Sigma(Q_0^2)   \Bigg[ \,  \frac{\Delta P^{(0)}_{Gq}}{\beta_0} 
+a_0\,  \frac{-\beta_1 \,\Delta P^{(0)}_{Gq} +\beta_0\, \Delta P^{(1)}_{Gq}  +\Delta P^{(0)}_{Gq} \,\Delta P_{\Sigma\Sigma}^{(1)}}{2\beta_0^2}
 \nn \\[2mm]
&+& \frac{a_0^2}{6\beta_0^3}  \Bigg( 2 \beta_1^2 \,\Delta P^{(0)}_{Gq}  - 2 \beta_0 \,\beta_2 \,\Delta P^{(0)}_{Gq}  - 
2 \beta_0 \,\beta_1 \,\Delta P^{(1)}_{Gq}
 + 2 \beta_0^2 \,\Delta P^{(2)}_{Gq}  \nn  \\[2mm]
&+& 2 \beta_0  \,\Delta P^{(0)}_{Gq} \,\Delta P_{\Sigma\Sigma}^{(2)} -
3 \beta_1 \,\Delta P^{(0)}_{Gq}  \, \Delta P_{\Sigma\Sigma}^{(1)}  +  \beta_0 \,\Delta P^{(1)}_{Gq}  \,\Delta P_{\Sigma\Sigma}^{(1)}
+ \Delta P^{(0)}_{Gq}  \, \big(\Delta P_{\Sigma\Sigma}^{(1)} \big)^2 \Bigg)\,
 \Bigg] \nn \\[2mm]
&=& - \left[ \, \frac{4}{9} + \frac{166}{81} a_0 - \left(\, \frac{7561}{2187} -\frac{160}{9} \zeta_3 \right)
a_0^2\,\right] \Delta\Sigma(Q_0^2) \, \simeq \,-0.125\,,
\eeeq
 with an asymptotic value given by
\beeq
 \label{staticnloinf1}
\Delta G_{\mathrm{stat}}^{{\mathrm{NNLO}}}(\infty) &=& -  \Delta\Sigma(Q_0^2)  \left[  \,\frac{\Delta P^{(0)}_{Gq}}{\beta_0} 
+a_0  \frac{ \Delta P^{(0)}_{Gq} \,\Delta P_{\Sigma\Sigma}^{(1)}}{\beta_0^2}\right. \nn\\[2mm]
&&\left. \hspace*{2.15cm}+\,a_0^2   \, \Delta P^{(0)}_{Gq}   \;
 \frac{ -\beta_1 \,\Delta P_{\Sigma\Sigma}^{(1)}+ \big(\Delta P_{\Sigma\Sigma}^{(1)}\big)^2 + \beta_0 \, \Delta P_{\Sigma\Sigma}^{(2)}
 }{2 \beta_0^3}\,\right] \nn \\[2mm]
&=& - \left[ \,\frac{4}{9} - \frac{32}{27}\, a_0 -\frac{1328}{243}\,a_0^2\right] \Delta\Sigma(Q_0^2)  \,\simeq \,-0.102\,.
\eeeq

\begin{figure}[h]
\vspace{4mm}
 \centerline{  \epsfig{figure=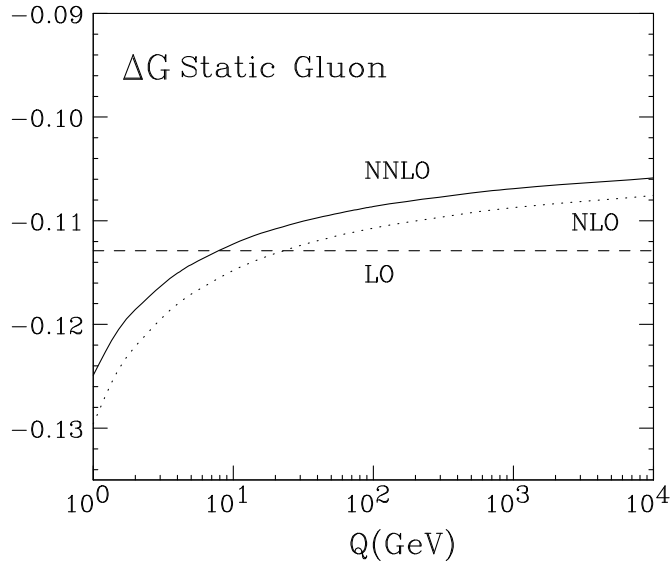,width=0.55\textwidth,clip=}}
\caption{ \label{fig:glue} Evolution of the "static" gluon solutions, starting from the inital scale $Q_0=1$ GeV.}
\end{figure}

Numerical results for the ``static'' solutions for $\Delta G$ are shown in Fig.~\ref{fig:glue}.
%%%%%%%%%%%%%%%%%%%%%%%%%%%%%%%%%%%%%%%%%%%%%%%%%%%
%%%%%%%%%%%%%%%%%%%%%%%%%%%%%%%%%%%%%%%%%%%%%%%%%%%    

 At NNLO the sum of the contributions 
by quarks and gluons starts at $\Delta\Sigma(Q_0^2)/2+G_{\mathrm{stat}}^{{\mathrm{NNLO}}}(Q_0^2) =0.0025$ with an asymptotic result of 
$\Delta\Sigma(\infty)/2+G_{\mathrm{stat}}^{{\mathrm{NNLO}}}(\infty) =0.01335$. In that particular scenario the total
orbital angular momentum almost accounts for the entire proton spin, $L_q+L_g\simeq 1/2$, 
and is almost constant in $Q^2$ (from $Q_0^2=1$ to $\infty$ it varies by less than 3\%).

As mentioned above, in our study of the ``static'' $\Delta G$ we have for simplicity chosen a fixed number of flavors, $N_f=3$. 
This will not be entirely adequate when considering the limit of large $Q^2$, and a matching to $N_f=4$ and $N_f=5$
should be performed at the charm and bottom mass scales, respectively. For the inputs 
$\Delta G_{\mathrm{stat}}(Q_0^2)$ given explicitly above, each matching would slightly upset the 
cancelation of the $1/a_s$ term in the solution for $\Delta G$, so that the resulting gluon spin contribution 
would not be entirely ``static'' anymore. However, this is expected to be a small effect. We have checked
that using a fixed number of $N_f$ throughout changes the asymptotic value of the static $\Delta G$ by
less than $10\%$. In any case, it is clear that a static solution for $\Delta G$ exists even if one performs
a full matching to $N_f=4$ and $N_f=5$: One could always start with an input at the bottom mass, $Q_0^2=m_b^2$,
that creates a static solution for all higher $Q^2$ with $N_f=5$. This solution could then be evolved backward to 
any lower $Q^2$ one desires, even to $Q^2=1~{\mathrm{GeV}}^2$ where $N_f=3$. This result at $Q^2=1~{\mathrm{GeV}}^2$
would then be the input to be used to obtain a static solution with full matching. 

We believe these solutions, especially because of the fact that they have a well behaved asymptotic limit at large scales, deserve further attention since they arise as  strong boundaries on non-perturbative physics from almost purely perturbative considerations.

\section{Non-singlet evolution of the valence quark spin contribution}

We finally turn to the evolution in the non-singlet sector. As discussed in the Introduction, 
we focus here on the strangeness ``valence'' spin contribution $(\Delta s-\Delta \bar{s})(Q^2)$
generated by three-loop evolution. 

Each of the non-singlet evolution equations in Eq.~(\ref{eq:evolnons}) has the solution
\beq
\label{sol}
\Delta q^{(A)}(Q^2) = U^{(A)}(Q,Q_0) \;\Delta q^{(A)}(Q_0^2) \;\;, \quad (A=V,\pm)  \; ,
\eeq
where $\Delta q^{(A)}(Q_0^2)$ is the corresponding input non-singlet combination
and the evolution operator $U^{(A)}$ is given by
\beq
\label{evop}
U^{(A)}(Q,Q_0) \,=\, \exp \left\{ \int_{Q_0^2}^{Q^2} 
\frac{dq^2}{q^2} \;\Delta P^{(A)}(a_s(q^2)) \right\} \;\;.
\eeq
We may readily use~(\ref{sol}) with $A=-$ and $A=V$ to obtain the solution for a valence quark contribution 
$\Delta q- \Delta {\bar q}$, resulting in~\cite{Catani:2004nc}
\beq
\label{sasym}
\left( \Delta q - \Delta {\bar q} \right)(Q^2) \,=\, 
U^{(-)}(Q,Q_0) \left[ \,\left( \Delta q - \Delta {\bar q} \right)(Q_0^2)
+ \frac{1}{N_f} 
\left( \frac{U^{(V)}(Q,Q_0)}{U^{(-)}(Q,Q_0)} - 1
\right) \Delta q^{(V)}(Q_0^2) \right] \;,
\eeq
where $\Delta q^{(V)}(Q_0^2)= \sum_q \left( \Delta q - \Delta \bar{q} \,\right)(Q_0^2)$ is the {\it total}
spin-dependent valence distribution in the nucleon as defined in Eq.~(\ref{eq:defnons}), at the initial 
scale. The first term on the right represents the homogenous component of the evolution of the valence 
distribution, which starts at NLO; its explicit expression is identical to the one in 
Eq.~(\ref{eq:singsol}) with the change $\Delta P_{\Sigma\Sigma} \rightarrow \Delta P^{(-)}$.
As follows from Eq.~(\ref{PVpm}), $\Delta P^{(V)} - \Delta P^{(-)} = N_f \left( \Delta P_{qq}^S - \Delta P_{q\bar{q}}^S
\right)$, which (see Eqs.~(\ref{LOP})--(\ref{NNLOP})) becomes nonzero starting from NNLO.
Therefore, the second term on the right of~(\ref{sasym}) will in general be non-vanishing as well, as long as 
$\Delta q^{(V)}(Q_0^2)\neq 0$, which of course is the case. We conclude that NNLO evolution generates
an asymmetry $\Delta s\neq\Delta\bar{s}$ in the contribution by strange and anti-strange quarks 
to the proton spin, {\it even} if $\Delta s=\Delta\bar{s}$ at the initial scale $Q_0$. 
This is at variance with the spin-averaged case where the first moment of $s-\bar{s}$ is protected
by the fact that there can be no net valence strangeness in the proton and remains zero to all orders. 

To NNLO accuracy, the evolution factor in the second term in Eq.(\ref{sasym}) reduces to
\beq 
\label{Uratio}
\frac{U^{(V)}(Q,Q_0)}{U^{(-)}(Q,Q_0)} - 1 
\,=\, - \frac{N_f\left(\Delta P^{(2)S}_{qq}  - \Delta P^{(2)S}_{q\bar{q}}\,\right)}{2 \beta_0} \,
\left( a_Q^2-  a_0^2 \right) \;,
\eeq
where, as before $a_Q=a_s(Q^2)$ and $a_0=a_s(Q_0^2)$. Therefore, assuming 
$\Delta s(Q_0^2) = \Delta \bar{s}(Q_0^2)$ in order to estimate the purely perturbative effect, 
we have, to NNLO
\beeq 
\label{finasol}
&&\hspace*{-1cm}\left( \Delta s - \Delta {\bar s} \right)_{\mathrm{pert}}(Q^2) \,=\,
- \frac{\Delta P^{(2)S}_{qq}  - \Delta P^{(2)S}_{q\bar{q}}}{2 \beta_0} \,
\left( a_Q^2-  a_0^2 \right) \left( \Delta u-\Delta\bar{u} + \Delta d-\Delta\bar{d} \,  \right)(Q_0^2)\nn\\[2mm]
&&\hspace*{4.5mm}\,=\,-\frac{5(23-12\zeta_2-16\zeta_3)}{72\beta_0\pi^2}\left( \alpha_s(Q^2)-  \alpha_s(Q_0^2) \right)
\left( \Delta u-\Delta\bar{u} + \Delta d-\Delta\bar{d} \,  \right)(Q_0^2),
\eeeq
where in the second line we have inserted the explicit value of $\left(\Delta P^{(2)S}_{qq}  - \Delta P^{(2)S}_{q\bar{q}}\,\right)$
from Eq.~(\ref{NNLOP}). The last factor on the right is of course just the total valence spin contribution at the initial scale.

We estimate the polarized strange asymmetry generated perturbatively by assuming, for example, $Q_0=1$~GeV 
with $\alpha_s(Q_0^2)=0.404$, and~\cite{deFlorian:2009vb,deFlorian:2008mr,deFlorian:2014yva} 
$\left( \Delta u-\Delta\bar{u} + \Delta d-\Delta\bar{d} \,  \right)(Q_0^2) \sim 0.5$, for which
\beq
\left( \Delta s - \Delta {\bar s} \right)_{\mathrm{pert}}(Q^2=10\,{\mathrm{GeV}}^2)\,\approx\,-6\,\cdot\,10^{-4}\,.
\eeq
%%%%%%%%%%%%%%%%%%%%%%%%%%%%%%%%%%%%%%%%%%%%%%%%%%%
\begin{figure}[h]
\vspace*{6mm}
 \centerline{  \epsfig{figure=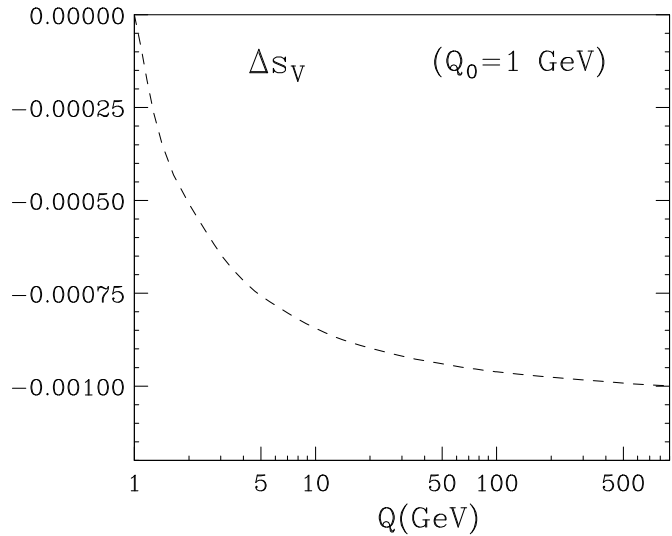,width=0.55\textwidth,clip=}}
\caption{ \label{asym} Evolution of the perturbatively generated ``valence'' strange 
contribution $\Delta s_{\mathrm{V}}\equiv  \Delta s - \Delta {\bar s}$, starting from the inital scale $Q_0=1$ GeV.}
\end{figure}                                            
%%%%%%%%%%%%%%%%%%%%%%%%%%%%%%%%%%%%%%%%%%%%%%%%%%%    
Figure \ref{asym} shows $\Delta s_{\mathrm{V}}\equiv  \Delta s - \Delta {\bar s}$ as a function of $Q$. 
The difference reaches $-0.001$ at $Q=M_Z$. As expected for a three-loop effect, it is small. On the other hand,
for the latest extractions of $(\Delta s+\Delta\bar{s})$~\cite{Ethier:2017zbq} the relative asymmetry
$|\Delta s-\Delta\bar{s}|/|\Delta s+\Delta\bar{s}|$ would be of order $1\%$. Evidently, non-perturbative
contributions~\cite{Wang:2016ndh} may well be the dominant source of the polarized strangeness
asymmetry. However, the effect we describe here would certainly need to be taken into
account in a full analysis. We emphasize that the perturbative asymmetry is robustly predicted to be 
negative, so that $\Delta \bar{s}>\Delta s$. 

\section{Conclusions}

We have presented a set of studies of the evolution of the quark and gluon spin contributions
to the proton spin at higher orders in perturbation theory, motivated by the 
recent calculations of the helicity splitting functions at full NNLO~\cite{Vogt:2008yw,Moch:2014sna,Moch:2015usa}.
We have argued that the evolution of $\Delta \Sigma$ is known even to four loops, which 
may prove valuable for lattice studies, as well as for comparisons to models residing at lower ``hadronic'' scales. 
The anomalous dimension relevant for the evolution of $\Delta\Sigma$ and related to the axial
anomaly also turns out to generate the evolution of $\Delta G$. The same must then be true
for the total orbital angular momentum $L_q+L_g$ in the Jaffe-Manohar sum rule, 
although the separate evolutions of $L_q$ and $L_g$ are presently only known to LO.

We have obtained analytical higher-order solutions for $\Delta \Sigma$ and $\Delta G$ and 
presented numerical results for their evolution. These show a stable upturn of 
$\Delta \Sigma$ toward low scales, bringing it actually closer to quark model expectations 
that favor a large quark spin contribution to the proton spin. The gluon spin $\Delta G$, 
when evolved backwards, shows a remarkable focus towards low values, again in line with quark model assumptions, 
setting a strong constraint on the gluon contribution at {\it hadronic} scales.
We have also shown that at every order of the perturbative evolution, there is a unique
solution for which $\Delta G$ tends to a finite asymptotic value as the scale becomes large. 
We have estimated the values of $\Delta G$ in such a scenario. 

We have finally also examined the size of the new effect arising from three-loop evolution in the
flavor non-singlet sector, the generation of an asymmetry in the strange and anti-strange
contributions to the proton spin. We have found that perturbative evolution predicts
$\Delta s - \Delta {\bar s}$ to be negative, with a magnitude of order $1\%$ of the total
$\Delta s +\Delta {\bar s}$. 
 
\section*{Acknowledgments}
We are grateful to Marco Stratmann for helpful discussions. The work of D.de F. has been partially 
supported by Conicet, ANPCyT and the Alexander von Humboldt Foundation.

\end{document}